\UseRawInputEncoding
\documentclass[aps,10pt,prl,twocolumn,superscriptaddress,noeprint,longbibliography,floatfix]{revtex4-2}
\PassOptionsToPackage{%
  pdfstartview=FitH,
  breaklinks=true,
  bookmarks=true,
  colorlinks=true,
  anchorcolor=black,
  citecolor=blue,
  filecolor=black,
  menucolor=black,
  urlcolor=blue,
  linkcolor=blue
}{hyperref}
\usepackage{graphicx}
\usepackage{bm}
\usepackage{amsmath, amssymb}
\usepackage{booktabs}
\usepackage{mathrsfs}
\usepackage{makecell}
\usepackage[normalem]{ulem}
\usepackage{pifont}
\usepackage{placeins}

\usepackage[mathscr]{euscript}
\usepackage{multirow}
\usepackage{physics}
\usepackage{xcolor}
\usepackage{siunitx}
\usepackage{placeins}
\usepackage{longtable}
\usepackage{orcidlink} 
\usepackage{titlesec}
\usepackage{nicefrac}
\usepackage{appendix}
\usepackage{mathtools}
\usepackage{tikz}
\usetikzlibrary{arrows.meta, positioning, calc, decorations.pathmorphing}
\usepackage{pgfplots}

\newcommand{\Heff}{H_{\rm eff}}

\newcommand{\Geff}{\Gamma_{\rm eff}}

\newcommand{\figref}[2]{\hyperref[#1]{\autoref*{#1}(#2)}}
\newcommand{\panels}[2]{\hyperref[#1]{panels~(#2)}}
\newcommand{\Panels}[2]{\hyperref[#1]{Panels~(#2)}}

\begin{document}
\title{Dissipative Stabilization of Floquet-Engineered Many-Body Order}

\author{Shreyas Raman \orcidlink{0000-0001-6640-9704}}
\thanks{equal contribution; \href{mailto:shreyasr@bu.edu}     {shreyasr@bu.edu}}
\affiliation{Department of Physics, Boston University, Boston, Massachusetts 02215, USA}
\author{Robin Sch\"afer \orcidlink{0000-0001-9728-2371}}
\thanks{equal contribution; \href{mailto:robin_schaefer@fas.harvard.edu}     {robin\_schaefer@fas.harvard.edu}}
\affiliation{Department of Physics, Harvard University, Cambridge, MA 02138, USA}

\author{Alicia J.  Koll\'ar}
\affiliation{Department of Physics, University of Maryland, College Park, MD 20742, USA}
\affiliation{Joint Quantum Institute, NIST/University of Maryland, College Park, Maryland 20742 USA}
\affiliation{Maryland Quantum Materials Center, Department of Physics, University of Maryland, College Park, MD 20742, USA}

\author{Anushya Chandran}
\affiliation{Department of Physics, Boston University, Boston, Massachusetts 02215, USA}

\date{\today}
\begin{abstract}
Floquet driving underlies Hamiltonian and gate engineering, and produces dynamical orders with no equilibrium counterpart. These phenomena are, however, transient in well-isolated systems. We show that statically coupled dissipative auxiliaries cool the system toward low-energy states of the Floquet-engineered Hamiltonian, stabilizing its low-temperature orders in the steady state. The excess energy density is controllably small at high drive frequency and weak coupling and is captured by a rate theory
based on Fermi's Golden Rule. Numerics on a Floquet-engineered transverse-field Ising chain confirm the rate theory on both sides of its ground-state transition. In a long-range Ising chain, which orders at finite temperature, the steady state is a discrete time crystal with period-doubled magnetization for appropriate drives. Our results
provide a rare analytical handle on the steady states of driven dissipative systems.
\end{abstract}
\maketitle

Floquet engineering uses strong and high-frequency periodic drives to realize effective many-body Hamiltonians~\cite{bukov_universal_high_2015, goldman_periodically_driven_2014, oka_floquet_engineering_2019, eckardt_atomic_quantum_2017, Rudner:2020aa, rahav_effective_hamiltonians_2003, Ozawa:2019Review}. High  frequency suppresses resonant transitions, but significantly dresses the time-averaged Hamiltonian~\cite{kuwahara_floquet_magnus_2016, abanin_effective_hamiltonians_2017}.
These effective Hamiltonians form the bedrock of quantum computing and simulation: to engineer qubits and gate operations~\cite{Huang:2021,Wang:2021FloquetGates}, to realize synthetic gauge fields and topological band structures~\cite{eckardt_atomic_quantum_2017,Weitenberg:2021aa,Rudner:2020aa, Ozawa:2019Review}, and to realize exotic dynamical orders that are impossible in thermal equilibrium, such as discrete time crystals~\cite{Zaletel:2023RMP,khemani2019briefhistorytimecrystals, choi_observation_discrete_2017, zhang_observation_discrete_2017, mi_time_crystalline_2022, else_prethermal_phases_2017, Gong_2018, Gambetta_2019, Ke_ler_2021} and anomalous Floquet topological phases~\cite{Rudner:2020aa,Wintersperger:2020aa, Roy:2017aa}.

Unluckily for the quantum mechanic, Floquet-engineered states are generically prethermal~\cite{abanin_exponentially_slow_2015, mori_rigorous_bound_2016, else_prethermal_phases_2017, Ho:2023aa, Machado_2020}. Generic isolated many-body systems absorb energy from the drive and approach a featureless infinite temperature state with no order at late times~\cite{dalessio_long_time_2014, lazarides_equilibrium_states_2014, Ponte:2015zr, Rubio:2020aa}.
Although this time is exponentially long at high frequency, this and other forms of heating limit driven protocols for state preparation and quantum simulation.
A big open problem in Floquet engineering is thus how to combat heating.

\begin{figure}[h!]
    \centering
    \includegraphics[width=\linewidth]{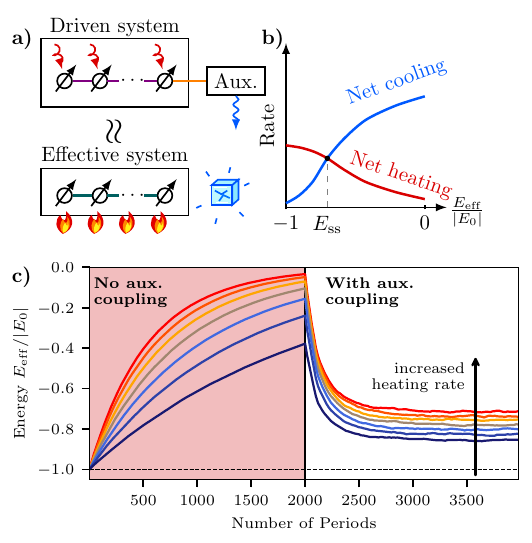}
    \caption{(a) A driven system coupled to a dissipative auxiliary maps to an effective static system subject to heating and cooling processes. (b) The steady-state effective energy $E_{\rm ss}$ is set by the balance between net heating and net cooling rates. (c) Normalized effective energy \(E_{\rm eff}/|E_0|\) for the driven Ising chain in~\autoref{eq:drive} with period-jitter noise and $L=18$. The system starts in the ground state of $\Heff$ with energy $E_0$, and the auxiliary coupling is switched on after 2000 periods. Without the auxiliary, the system heats toward infinite temperature; with the auxiliary, the system relaxes to a low-energy steady state, whose energy varies with the heating rate.}
    \label{fig:1}
\end{figure}

One strategy for static Hamiltonians is dissipative cooling, wherein an auxiliary absorbs excitations and releases energy into the environment, driving the system toward its ground-state manifold~\cite{poyatosQuantumReservoirEngineering1996, murch_cavity_assisted_2012, hacohengourgy_cooling_autonomous_2015, maDissipativelyStabilizedMott2019}. Floquet systems pose two obstacles for this kind of approach. First, cooling must continuously compete with heating from the drive~\cite{Tsuji_2009, Dehghani_2014, Iwahori_2017, Ikeda_2020} on top of external noise, and from the auxiliary itself~\cite{seetharamControlledPopulationFloquetBloch2015, iadecolaOccupationTopologicalFloquet2015, moriFloquetStatesOpen2023}. Second, Floquet quasi-energies are defined only modulo the drive frequency, so there is no lowest state to target. At high frequency, the effective Hamiltonian $\Heff$ generating time evolution unfolds the quasi-energy spectrum and defines ground states, a feature recent works have begun to exploit for cooling Floquet systems~\cite{shiraiEffectiveFloquetGibbsStates2016, petiziol_cavity_based_2022, miStableQuantumCorrelatedMany2024, schnellDissipativePreparationFloquet2024, wanckel_dissipative_floquet_2026, steinfadt_dissipation_assisted_2026, diermannFloquetstateCooling2019}. Yet even with a target state, it remains open whether cooling can generically overcome heating and stabilize the ordered phases targeted by Floquet engineering.

We demonstrate the stabilization of Floquet-engineered order in the steady state using experimentally feasible, statically coupled, finite-bandwidth dissipative auxiliaries~\cite{ritter_autonomous_stabilization_2025} that resonantly absorb excitations of $\Heff$ [\autoref{fig:1}(a)]. We develop a general theory of cooling in this framework, and derive the heating and cooling rates at weak coupling. Their balance determines the steady-state energy density [\autoref{fig:1}(b)] and identifies optimal auxiliary parameters. \autoref{fig:1}(c) illustrates this for a driven spin chain with an auxiliary at one end: without it, heating drives the effective energy toward its infinite-temperature value, while a well-matched auxiliary rapidly cools the system to a steady state close to the ground state of $\Heff$. With a finite density of auxiliaries, the excess energy density is controllably small at weak coupling, and the steady state inherits the low-temperature order of $\Heff$; we find discrete time-crystalline order in the steady state of a long-range Ising chain.

\paragraph{Analytic framework. }
Consider a system driven with period $T=2\pi/\omega$, with Hamiltonian $H(t+T)=H(t)$. A high-frequency Schrieffer-Wolff transformation yields a rotating frame governed by a static effective Hamiltonian up to order $n$ in $1/\omega$~\cite{bukov_universal_high_2015}:
\begin{equation}
    K(t)H(t)K^\dagger(t) + i\dot K(t)K^\dagger(t) = \Heff + \mathcal{O}\left(\omega^{-(n+1)}\right),
    \label{eq:Hrot}
\end{equation}
where $K(t)$ is unitary~\cite{shavitt_quasidegenerate_perturbation_1980}.
For local many-body systems, the expansion is asymptotic but controlled up to an optimal order $n_*\sim\omega/\Omega$, where $\Omega$ is a local energy scale of the system~\cite{kuwahara_floquet_magnus_2016, mori_rigorous_bound_2016, abanin_exponentially_slow_2015}.
The residual time-dependent terms at orders $\geq n+1$ produce transitions between eigenstates of $\Heff$ and a slow increase of the effective energy $E_{\rm eff}(t)=\mathrm{Tr}[K(t)\rho(t)K^\dagger(t)\Heff]$~\cite{abanin_effective_hamiltonians_2017}, where $\rho(t)$ is the lab frame density matrix of the system at time $t$. This heating is intrinsic: it is present even without the auxiliary.

The system is coupled to $N_{\rm aux}$ dissipative auxiliaries, each being the simplest finite-bandwidth element available: a spin-1/2. Each auxiliary spin has splitting $H_{\rm aux}=\frac{\Delta}{2}\tau^z$, coupling $H_{\rm c}=\frac{g}{4}\tau^y O$ to a local system operator $O$, and $T_1$ decay rate $\kappa$, modeled by the jump operator $L_{\rm aux}=\sqrt{\kappa}\tau^-$.

Each auxiliary cools the system because it preferentially decays from $\ket{\uparrow}$ to $\ket{\downarrow}$. When $\kappa \ll \Delta$, this asymmetry makes transitions that remove 
energy from the system far more probable than those that add energy to it. 
In the rotating frame, the system-auxiliary coupling is time-dependent; 
writing $O_{\rm rot}(t) = K(t)OK^\dagger(t) = \sum_m O^{(m)}e^{-im\omega t}$, 
the (most important) $m=0$ term resonantly targets transitions $\ket{\alpha,\downarrow}\to
\ket{\beta,\uparrow}$ with $E_\alpha - E_\beta \approx \Delta$, so that the auxiliary selectively absorbs excitations near its own energy scale. To cool to the ground state, $\Delta$ is matched with the intrinsic low energy scales of $\Heff$. 

Since energy-absorbing off-resonant transitions are allowed in the system at a small rate, the auxiliaries act as a narrow-band cold bath for the system, and not an ideal zero-temperature bath.

For $m\geq 1$ and $\omega>\Delta$, the $m^{\rm th}$ harmonic of the coupling can resonantly excite the auxiliary while depositing energy $m\omega-\Delta$ in the system. These sideband processes have been discussed before, see e.g. Ref.~\cite{petiziol_cavity_based_2022}. Their rates are exponentially small when $\kappa \ll \Delta, \omega$, that is, for a narrow-band cold bath, because of the small spectral weight of $O^{(m)}$ at the energy difference $m\omega-\Delta$. If the system were instead connected to a broadband zero temperature Ohmic bath, these processes would compete with the resonant $m=0$ process that dominantly cools~\cite{iadecolaOccupationTopologicalFloquet2015}.

\begin{figure*}[t]
\centering
\includegraphics[width=\linewidth]{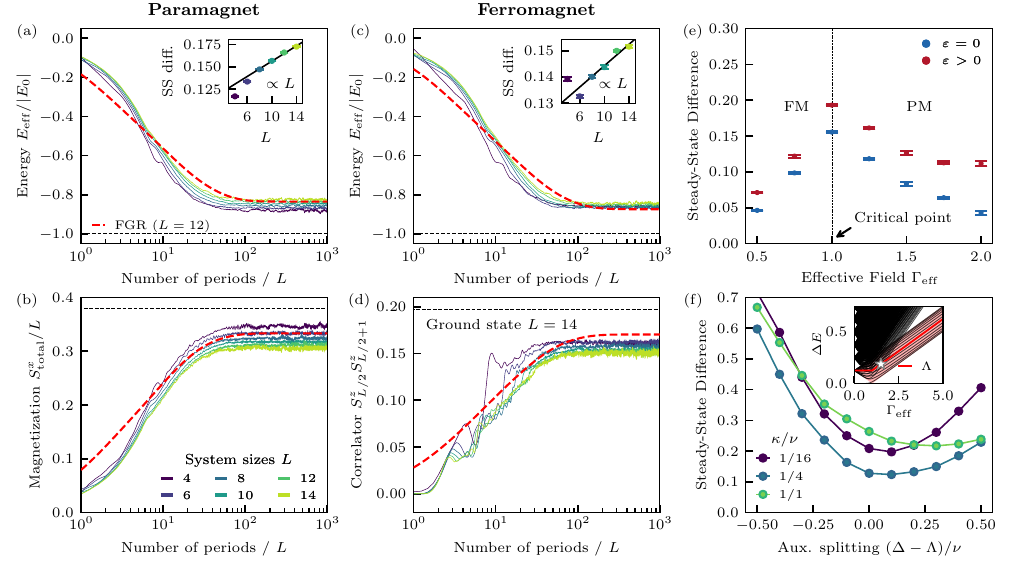}
\caption{Cooling into low-energy states of a Floquet-engineered
transverse-field Ising chain using a single auxiliary coupled at one end.
(a),(b)
Paramagnetic ($\Gamma_{\rm eff}=1.5$) and (c),(d) ferromagnetic
($\Gamma_{\rm eff}=0.75$)  regimes, starting from the infinite-temperature
state, versus rescaled time. (a),(c) Normalized effective energy; red dashed curves in (a)--(d) show the FGR rate-equation prediction~[\autoref{eq:pauli}]; black
dotted lines mark ground-state values. Insets: steady-state energy
difference $\frac{E_{\rm ss}-E_0}{|E_0|}$ vs.\ $L$.
(b),(d) Transverse magnetization and
nearest-neighbor Ising correlator, respectively. (e) Steady-state energy
difference vs.\ $\Gamma_{\rm eff}$ after $10\, 000$ periods for $L=12$, with ($\varepsilon>0$) and
without ($\varepsilon=0$) period jitter;
$\varepsilon$ is chosen to keep the net heating rate approximately fixed (see End Matter).
The black
dashed line marks the critical point at zero temperature. For panels (a)--(e), the parameters are \(\omega = 8\), \(\Delta=\Lambda\), \(4\kappa=g = \nu\). (f) Steady-state energy difference vs.\ rescaled auxiliary splitting after $12\, 000$ periods for several auxiliary decay rates $\kappa$. Here $L=12$, $\omega=8$, $\Gamma_{\rm eff}=1.5$, and $g=\nu$ for all linewidths. The jitter strength is the same as in panel (a), $\varepsilon\approx0.149$.
Inset: effective
spectrum $\Delta E = E_{\rm eff}-E_0$ vs.\ $\Gamma_{\rm eff}$, with the
single-quasiparticle band center shown in red;
the star marks the
parameter point used in panel (f).}
\label{fig:2}
\end{figure*}

External noise or imperfections in the drive can also heat the system. Finite Floquet systems do not heat at drive frequencies comparable to the many-body bandwidth, because resonant many-body transitions are sparse or absent~\cite{bukov_heating_many_2016}. In the numerics, which are discussed in the next section, to ensure thermalization at finite size, we add tunable and experimentally realistic noise: period jitter, drawing $\zeta_n$ uniformly from $[-\varepsilon, \varepsilon]$ and setting the period of the $n^{\rm th}$ drive cycle to be $T_n = T(1+\zeta_n)$~\cite{fleckenstein_prethermalization_thermalization_2021}. 

At large $\omega$, small $g$, and small $\varepsilon$, drive-, auxiliary-, and noise-induced processes act independently, and the eigenstate populations $p_\alpha=\bra{\alpha}K\rho K^\dagger\ket{\alpha}$ in the basis of $\Heff$ obey a rate equation,
\begin{equation}
    \dot p_\alpha = \sum_\beta\left[W_{\beta\to\alpha}p_\beta - W_{\alpha\to\beta}p_\alpha\right],
    \label{eq:pauli}
\end{equation}
where $W_{\alpha \to \beta} = W_{\alpha \to \beta}^{\rm int} + W_{\alpha \to \beta}^{\rm noise} + W_{\alpha \to \beta}^{\rm aux}$ is the total transition rate from $\ket \alpha \to \ket \beta$, and $W_{\alpha \to \beta}^{\rm int/noise/aux}$ is the transition rate induced by intrinsic heating processes/external noise/coupling to the auxiliary. Expressions for the perturbative transition rates obtained by Fermi's Golden Rule (FGR) are given in the End Matter.

The residual terms in \autoref{eq:Hrot}, external noise, auxiliary sideband processes, and off-resonant auxiliary processes contribute to the net heating rate. Resonant cooling by the auxiliary sets the net cooling rate. Equality of the net heating and cooling rates determines the energy of the steady state, as estimated in ~\autoref{eq:e_ss} and shown in \autoref{fig:1}(b). 

Optimal auxiliary parameters for cooling can be derived in the perturbative regime. Defining the low-energy excitation band of $\Heff$ to have center $\Lambda$ and bandwidth $\nu$, we require,
\begin{align}
   (\kappa \sim \nu) \lesssim (\Delta\sim \Lambda), \quad \frac{g}{4} \lesssim \kappa, \quad \Omega,\Delta,\kappa,g \ll \omega. \label{Eq:GoodRegime}
    \end{align}
The first condition ensures that the entire low-energy excitation band can be resonantly targeted by the auxiliary, and that the auxiliary preferentially decays to $|\downarrow \rangle$.
The second condition prevents the auxiliary from strongly hybridizing with the system and thus being driven out of resonance with the low-energy band.
The third condition ensures that the dominant unitary dynamics is generated by \(\Heff\) over a long prethermal window, within which the steady state is reached, and that sideband auxiliary heating processes are suppressed.

The steady state energy difference in the perturbative regime, up to model-dependent spectral details, is (see End Matter for a derivation): 
\begin{equation}
    \frac{E_{\rm ss} \!-\! E_0}{|E_0|}
    \!\sim\!
    \frac{\kappa \omega}{n_{\rm aux}g^2}\!\left[\!\left(\!1 \!+\! \frac{n_{\rm aux} g^2}{\Delta^2}\!\right)\!e^{-\frac{\omega}{\Delta}} \!+\! \varepsilon^2\frac{\Delta^2}{\omega^2}\! \right] \!+\! \frac{\kappa^2}{\Delta^2},
    \label{eq:e_ss}
\end{equation}
where $E_0$ is the ground-state energy of $\Heff$ and $n_{\rm aux} = N_{\rm aux}/L$ is the number density of auxiliaries. Thus, the steady state is close to the ground state in energy density when the right-hand side of \autoref{eq:e_ss} is small.

\paragraph{Numerical tests in a driven Ising model. }
We demonstrate dissipative cooling in a Floquet-engineered Ising chain, whose effective Hamiltonian is a nearest neighbor Transverse Field Ising model (NN-TFIM), of up to $L=18$ spins coupled to $N_{\rm aux}=1$ auxiliary. 

The NN-TFIM has no order at finite temperature. It is nevertheless a useful benchmark for cooling because (i) it is exactly soluble, and (ii) there are vestiges of the ground-state order in the steady state at accessible $L$. A single auxiliary permits larger systems in a proof-of-principle demonstration but cannot cool to a low $E_{ss}/L$ as $L \to \infty$;
that requires a finite density of auxiliaries (see the Supplemental Material (SM) for evidence that $N_{\rm aux} = 2$ cools twice as fast as $N_{\rm aux} = 1$). The chain is subject to a four-step drive,
\begin{equation}
    H(t) = H_0 + f_1(t)\frac{B}{2}H_y + f_2(t)\frac{B}{2}H_z,
    \label{eq:drive}
\end{equation}
with $H_0=-\frac{J}{4}\sum_i\sigma_i^z\sigma_{i+1}^z$, $H_\alpha=\sum_i\sigma_i^\alpha$, and $(f_1,f_2)=(1,0),(0,1),(-1,0),(0,-1)$ on successive quarter periods. We set $J = 1$ going forward.

To order $1/\omega$, $\Heff = H_0 - \frac{B^2\pi}{16\omega}H_x$ with dimensionless transverse field $\Geff=\pi B^2/(4\omega J)$. The zero-temperature transition at $\Geff=1$ separates ferromagnetic ($\Geff<1$) and paramagnetic ($\Geff>1$) ground states. The quasiparticle excitation band above the ground state is centered at $\Lambda=\frac{J}{2}\max(1,\Geff)$ with bandwidth $\nu=J\min(1,\Geff)$. Higher-order terms in the drive expansion weakly break the integrability of the TFIM. They are included in the numerical simulations, which use the full time-dependent Hamiltonian \(H(t)\).

The auxiliary is coupled to the edge of the chain, and we choose $O=\sigma_1^y$, which can effectively remove quasiparticle excitations in both phases. We add period jitter, which contributes to net heating of the system as described in the previous section. Numerical methods are discussed further in the End Matter.

\autoref{fig:2} demonstrates cooling across the effective Ising phase diagram, shows quantitative agreement with the predictions of the FGR rate theory, and tests the auxiliary design rules. The first column corresponds to the paramagnetic regime, $\Gamma_{\rm eff}=1.5$, and the second to the
ferromagnetic regime, $\Gamma_{\rm eff}=0.75$. In~\figref{fig:2}{a-d}, time is plotted as the number of drive periods divided by $L$; the collapse of different system sizes on this rescaled axis shows that the cooling time grows approximately linearly with $L$, as expected when one local auxiliary cools an extensive chain.

Starting from states at infinite temperature with $E_{\rm eff}=0$, the
system relaxes to a low-energy steady state on a timescale $\sim
10^2 L$ drive periods~[\figref{fig:2}{a,c}]. The dashed red curves show the FGR rate-theory prediction, obtained by diagonalizing $H_{\rm eff}$ and computing the relevant rates as described in the End Matter; the dashed curves reliably capture both the cooling timescale and the steady-state energy. The insets show that the steady-state energy offset grows approximately linearly with system size, consistent with bulk heating and cooling by a single auxiliary.

The steady state also develops local magnetization (correlations) that are characteristic of the paramagnetic (ferromagnetic) effective
ground state [\figref{fig:2}{b,d}]. Note that the steady state is not expected to have long range order as $L \to \infty$. We probe physical observables in the lab frame, not the dressed operator corresponding to the order parameter in the rotating frame, i.e. for the observable $M$, $\langle 
M(t) \rangle = \rm Tr[\rho(t) M]$.

Dissipative cooling is effective across the phase diagram~[\figref{fig:2}{e}],
both without ($\varepsilon=0$) and with ($\varepsilon>0$) period jitter, with
the steady state energy difference (LHS of Eq.~\eqref{eq:e_ss}) being maximal near the ground state phase transition. There, the vanishing spectral gap stymies dissipative
cooling. \figref{fig:2}{f} confirms that the FGR rate theory correctly predicts optimal auxiliary
parameters: the steady-state energy difference is minimal when
$\Delta \simeq \Lambda$ and $\kappa \sim \nu$, and the broad minimum in
both shows that the cooling scheme does not require fine tuning. Simulation parameters are given in the caption of~\autoref{fig:2} and in the End Matter. 

\paragraph{Stabilizing a period-doubled response. }
We now turn to a driven and dissipatively cooled long-range Ising chain, whose steady state exhibits discrete time-crystalline order. Without cooling, the order is prethermal, so that the period-doubled magnetization response of initially polarized states decays to zero. The dissipative auxiliary cools the chain into period-two ferromagnetically ordered states; the accompanying period-doubled response is revealed by a weak period-two jitter.

\begin{figure}[t]
\centering
\includegraphics[width=\columnwidth]{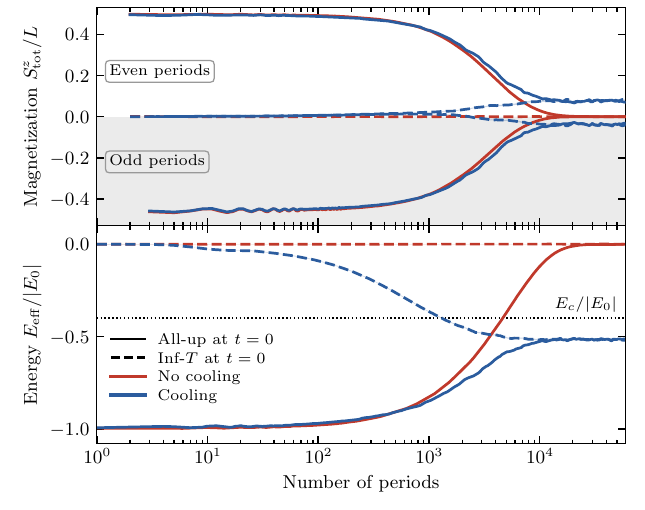}
\caption{Dissipative stabilization of a period-doubled magnetization response for Eq.~\eqref{Eq:Upipulse}. Upper panel: Stroboscopic magnetization on even (white) and odd (gray) cycles. Lower panel: Normalized effective energy; gray dotted line marks the critical value $E_c/|E_0|$. Red (blue) curve corresponds to no (attached) auxiliary, while the solid (dashed) line represents the initial state with all spins up (at infinite-temperature). Parameters are $L=12$, $p=1.5$, $\omega/J=16$, $B/J=2.257$, $\varepsilon_1=0.05$, $\varepsilon_2=0.07$, $\Delta/J=1$, $g/J=1$, and $\kappa/J=0.25$. The auxiliary couples to the first spin through $O=\sigma_1^y$.}
\label{fig:tc}

\end{figure}

Consider a drive protocol motivated by the experiments of Kyprianidis et al.~\cite{kyprianidis_observation_prethermal_2021}, with a one-period Floquet unitary given by,
\begin{equation}
    U(T)=R_xU_z(T/2)U_y(T/2),\label{Eq:Upipulse}
\end{equation}
where \(R_x=\prod_i\sigma_i^x\) is a global \(\pi\)-pulse, \(U_\alpha(t)=e^{-iH_\alpha t}\), $H_\alpha = H_0 + \frac{B}{2}\sum_i \sigma_i^\alpha$, and \(H_0 = -\frac{J}{4}\sum_{i>j}\frac{\sigma_i^z \sigma_j^z}{|i-j|^p}\). We use the interaction exponent $p=1.5$. Since conjugating by \(R_x\) flips the sign of \(\sigma^{y,z}\) and leaves \(H_0\) invariant, \(R_xH_\alpha R_x\equiv \tilde H_\alpha=H_0-\frac{B}{2}\sum_i \sigma_i^\alpha\). The two-period unitary is therefore
\begin{equation}
    U(2T)
    =
    \tilde U_z(T/2)\tilde U_y(T/2)
    U_z(T/2)U_y(T/2),
    \label{eq:two_period_tc}
\end{equation}
where \(\tilde U_\alpha(t)=e^{-i\tilde H_\alpha t}\). At high frequency, the effective Hamiltonian $\Heff$ corresponding to \autoref{eq:two_period_tc} is a long-range TFIM with ferromagnetic order below a critical energy density $E_c/L$. The ferromagnetic states ~\cite{dyson_existence_phasetransition_1969} exhibit a period-doubled response because \(R_x\) flips the total $z$-magnetization, $S^z_{\rm tot}$, after every period $T$. 

In the absence of cooling, the period-doubled response is prethermal, and the effective energy reaches that of the state at infinite temperature at late times, as shown in the red curves in~\autoref{fig:tc}. As before, period jitter of strength $\varepsilon_1$ ensures thermalization at finite size.

In the presence of cooling, the time-crystal phase becomes stable, even in the steady state. The auxiliary cools the chain below the equilibrium ordering threshold such that $E_{\rm eff} < E_{\mathrm{c}}$ (blue curves in the bottom panel of~\autoref{fig:tc}), allowing ferromagnetic correlations to survive in the steady state (see SM). However, with period-one jitter ($\varepsilon_1$) alone, the two ferromagnetic sectors are equally populated and the net magnetization vanishes.

Steady-state period-doubled magnetization is revealed by incorporating an independent period-two jitter component of strength $\varepsilon_2$, drawing one value for each pair of consecutive periods. The resulting heating rates differ between the two ferromagnetic sectors (see SM). Sector-independent cooling then favors the slower-heating sector, producing a steady-state period-doubled magnetization response independent of initial state, as shown in the top panel of~\autoref{fig:tc}.

\paragraph{Discussion. } 
We have shown that many-body Floquet systems can be cooled, and kept cold despite intrinsic heating, by a finite density of finite-bandwidth dissipative auxiliaries statically coupled to the system. Numerical simulations of driven Ising chains coupled to one auxiliary provide a proof of principle.
The scheme is simple, autonomous, and broadly applicable. 
Most strikingly, the steady state can
support discrete time-crystalline order. Cooling restores ferromagnetic order in $\Heff$ defined over $2T$,
and a weak period-two jitter selects one sector, yielding a stable period-doubled response from any initial state.

FGR rates for heating and cooling processes, valid in the limit of weak coupling to the local auxiliaries, show the existence of a hierarchy of energy scales [\autoref{Eq:GoodRegime}] in which the residual energy density in the steady state is controllably small.
The scheme cools generic Floquet systems with gapped effective ground states and quasiparticle excitations that can be efficiently removed by local coupling to auxiliaries. The required ingredients -- statically coupled, lossy qubits or cavities -- are readily available in superconducting-circuit, cavity-QED, trapped-ion, and in neutral-atom systems.
Combining this framework with non-local coupling to the auxiliary may provide a route to efficient state preparation and error correction in topological Floquet systems~\cite{dennis_topological_quantum_2002, hastings_dynamically_generated_2021, polla_quantum_digital_2021}.

\begin{acknowledgments}
The authors thank Paul Schindler, Marin Bukov, David Long, Sarang Gopalakrishnan, Fabian Essler, Samarth Hawaldar and Stewart Morawetz for helpful discussions.
This project was supported by AFOSR Grant No. FA9550-24-1-0121. R.S.
acknowledges support from the DFG under Project No. 575641691 and the Helmholtz-Zentrum Berlin. Simulations used DanceQ~\cite{schaefer_danceq_high_2025, schaefer_codebase_release_2025}.
\end{acknowledgments}

\renewcommand{\selectlanguage}[1]{} 
\bibliography{ref}

\newpage
\onecolumngrid
\vspace{1.5em}
\begin{center}
  \textbf{\large End Matter}
\end{center}
\vspace{0.5em}
\twocolumngrid

\paragraph{FGR Transition Rates and Steady-State Energy. }
Let $\Heff \ket \alpha = E_\alpha \ket \alpha$, and define $\delta E_{\beta \alpha} = E_\beta - E_\alpha$. In the weak-coupling and high-frequency regime discussed in the main text, intrinsic Floquet heating, heating from period jitter and all auxiliary-induced processes can be treated perturbatively, and the corresponding transition rates derived using FGR. 

The transition rates due to auxiliary-induced processes follow by treating $H_c = \frac{g}{4} \tau^y O$ perturbatively and tracing over the dissipative auxiliary~\cite{petiziol_cavity_based_2022}. Writing $O_{\rm rot}(t) = K(t) O K^\dagger(t) = \sum_m O^{(m)} e^{-im\omega t}$, one obtains 
\begin{multline}
    W_{\alpha \to \beta}^{\rm aux} = \frac{g^2}{16} \sum_m \left|\bra{\beta} O^{(m)} \ket \alpha\right|^2 \\
    \times \frac{\kappa}{(E_\alpha - E_\beta + m\omega - \Delta)^2 + \frac{\kappa^2}{4}}.
    \label{eq:em_aux_full}
\end{multline}
For $N_{\rm aux}$ independent auxiliaries, one sums~\autoref{eq:em_aux_full} over the auxiliary indices with the associated coupling operators. The most important is the $m=0$ term, which causes resonant transitions in the system with $E_\alpha - E_\beta \approx \Delta$ with rate $\sim g^2/\kappa$. The Lorentzian in~\autoref{eq:em_aux_full} also allows off-resonant transitions with $E_\alpha < E_\beta$ with much smaller rates $\sim g^2 \kappa/\Delta^2$. For $m\geq 1$, the sideband is centered at $E_\beta-E_\alpha=m\omega-\Delta$. On resonance the rate for a single pair of levels is $g^2|\bra\beta O^{(m)}\ket\alpha|^2/(4\kappa)$. Summing over final states with a spectral density smooth on the scale $\kappa$ replaces the Lorentzian peak by its integrated weight. Estimating the resulting spectral density of $O^{(m)}$ at the sideband energy as $\Omega^{-1}e^{-(m\omega-\Delta)/\Omega}$ gives a total rate of order $g^2\Omega^{-1}e^{-(m\omega-\Delta)/\Omega}$ per auxiliary.

Intrinsic Floquet heating with respect to $\Heff$ is generated by the residual terms after truncating the high-frequency expansion in~\autoref{eq:Hrot}. Calling these terms $V_{\rm res}(t) = \sum_m V_{\rm res}^{(m)}e^{-im\omega t}$, one obtains~\cite{ikeda_fermis_golden_2021}
\begin{equation}
    W_{\alpha \to \beta}^{\rm int} = 2\pi \sum_m \left|\bra{\beta}V_{\rm res}^{(m)}\ket \alpha\right|^2 \delta(E_\alpha - E_\beta + m\omega).
    \label{eq:em_int}
\end{equation}
Note that the $m = 0$ term does not contribute to intrinsic heating. Truncating the expansion near the optimal order exponentially suppresses intrinsic heating, with rate $\sim \Omega e^{-\omega/\Omega}$~\cite{abanin_exponentially_slow_2015, mori_rigorous_bound_2016, odea_prethermal_stability_2024}. 
In realistic systems, the delta function in~\autoref{eq:em_int} is broadened by imprecision in the drive frequency or finite observation time. 

Period jitter provides a tunable heating process in finite systems. The period in cycle $n$ is $T_n = T(1+ \zeta_n)$, with $\overline{\zeta_n} = 0$ and $\overline{\zeta_n \zeta_m} = \varepsilon^2 \delta_{nm}/3$. To linear order in the jitter strength $\varepsilon$, the effective Hamiltonian for period $n$ is stochastic and given by $\Heff^{(n)} = \Heff + \zeta_n A_{\rm jit} + (\zeta_{n+1}-\zeta_n)B_{\rm jit}$, where $\Heff$ is defined in~\autoref{eq:Hrot},  $A_{\rm jit}$ arises from the rescaled duration of the $n^{\rm th}$ cycle, and $B_{\rm jit}$ arises from the mismatch of the rotating frames between consecutive periods. The full derivation of this stochastic effective Hamiltonian is given in the Supplemental Material. Averaging over independent jitter realizations gives 
\begin{equation}
    W_{\alpha \to \beta}^{\rm noise} = \frac{4\varepsilon^2}{3T} \left|\bra{\beta}M_{\rm jit}\ket{\alpha}\right|^2 \frac{\sin^2(\delta E_{\beta \alpha} T/2)}{\delta E_{\beta \alpha}^2},
    \label{eq:W_jit}
\end{equation}
where $\bra{\beta}M_{\rm jit}\ket{\alpha} = \bra{\beta}A_{\rm jit}\ket \alpha + (e^{-i \delta E_{\beta \alpha} T} - 1)\bra{\beta} B_{\rm jit}\ket \alpha$. From~\autoref{eq:W_jit}, uncorrelated jitter across periods acts as a broadband heating channel for $\Heff$ with rate $\sim \varepsilon^2\Lambda^2/\omega$. 

The steady-state energy $E_{\rm ss}$ is set by the competition of the processes described in~\autoref{eq:em_aux_full}--\autoref{eq:W_jit}. To estimate it, we weight
each transition rate by the energy transferred and set the resulting net energy absorption rate to zero. 

Near the ground state of $\Heff$, we assume that the excitations form a dilute quasiparticle gas with density
$n_{\rm qp}\sim(E_{\rm eff}-E_0)/(\Lambda L)$, where $\Lambda$ is the typical quasiparticle energy. When $g/4 \lesssim \kappa$, the auxiliary typically decays after absorbing an excitation from the system, and suppresses repeated exchange of excitations with the system. At low quasiparticle density, the probability of finding a quasiparticle
near an auxiliary is proportional to $n_{\rm qp}$, so the resonant cooling rate per auxiliary scales as $(g^2/\kappa)n_{\rm qp}$. Off-resonant upward transitions are also allowed by the Lorentzian
in \autoref{eq:em_aux_full}. These create excitations even when no quasiparticles are initially present, giving a leading heating rate
$\sim g^2\kappa/\Delta^2$ per auxiliary that is independent of $n_{\rm qp}$. All auxiliary-induced processes scale with $N_{\rm aux}$, whereas heating from period jitter and intrinsic Floquet processes scales with the system size $L$.

To convert these rates to energy absorption rates, the rates need to be multiplied by the typical energy exchanged by each process. Resonant cooling removes energy $\sim\Lambda$ per quasiparticle,
while off-resonant auxiliary heating and period jitter create excitations of typical energy $\sim\Lambda$. Intrinsic Floquet
heating is dominated by events that absorb energy $\omega$ from the drive, and auxiliary sideband heating deposits energy $m\omega-\Delta$. The resulting energy balance is
\begin{multline}
    \dot E_{\rm eff} \sim -N_{\rm aux} \frac{g^2}{\kappa}n_{\rm qp} \Lambda + N_{\rm aux} \frac{g^2 \kappa}{\Delta^2} \Lambda
    + L \Omega e^{-\omega/\Omega}\omega \\+ L\varepsilon^2 \frac{\Lambda^3}{\omega} + \sum_{m\geq 1}N_{\rm aux} \frac{g^2}{\Omega} e^{(\Delta - m\omega)/\Omega} (m \omega - \Delta)
\end{multline}
Setting $\dot E_{\rm eff} = 0$, we get 
\begin{multline}
    \frac{E_{\rm ss} - E_0}{L} \sim \frac{\kappa^2}{\Delta^2}\Lambda + \frac{\omega \Omega e^{-\omega/\Omega}\kappa}{n_{\rm aux}g^2} + \frac{\varepsilon^2 \Lambda^3 \kappa}{n_{\rm aux} g^2 \omega} \\+ \sum_{m \geq 1}\frac{\kappa} {\Omega} e^{(\Delta - m\omega)/\Omega} (m\omega - \Delta)
\end{multline}
Using $|E_0|\sim\Omega L$ and estimating
$\Omega\sim\Lambda\approx\Delta$ for an optimally tuned auxiliary,
we obtain the scaling form
\begin{multline}
\frac{E_{\rm ss} - E_0}{|E_0|} \sim \frac{\kappa \omega}{n_{\rm aux}g^2}\left(e^{-\omega / \Delta} + \varepsilon^2 \frac{\Delta^2}{\omega^2} \right) +\frac{\kappa^2}{\Delta^2} \\
+ \sum_{m \geq 1} \frac{\kappa}{\Delta^2} (m\omega - \Delta)e^{\frac{\Delta - m\omega}{\Delta}}
\end{multline}
Relative to the bulk intrinsic heating estimate in \autoref{eq:em_int}, the estimated heating power of harmonic $m$ is smaller by a factor of order $n_{\rm aux}(g/\Delta)^2 (m-\Delta/\omega)e^{[\Delta-(m-1)\omega]/\Delta}$. If we retain only the dominant $m = 1$ term, and use $\omega - \Delta \approx \omega$, we obtain \autoref{eq:e_ss} in the main text.

\paragraph{Numerical implementation.}
The microscopic driven-dissipative dynamics of the \emph{combined} system and
auxiliary is governed by the Lindblad equation~\cite{carmichael_open_systems_1993}
\begin{equation}
    \dot\rho
    =
    -i[H_{\rm tot}(t),\rho]
    +
    \mathcal D[L_{\rm aux}]\rho,
    \label{eq:em_lindblad}
\end{equation}
where $\rho$ is the joint density matrix, \(H_{\rm tot}(t)=H(t)+H_{\rm aux}+H_{\rm c}\), and
\(\mathcal D[L]\rho=L\rho L^\dagger-\frac12\{L^\dagger L,\rho\}\).
We evolve the full time-dependent Hamiltonian \(H(t)\), not the truncated
\(\Heff\), so that higher-order terms in the drive expansion are retained. In the case of the long-range Ising model (simulating~\autoref{Eq:Upipulse}), when the auxiliary is present, each cycle ends with the joint pulse $P=\tau^z\otimes R_x$, which also commutes with the coupling $\tau^y\sigma_1^y$.
\autoref{eq:em_lindblad} is simulated using stochastic unraveling
of quantum trajectories for chains of up to \(L=18\) spins~\cite{carmichael_open_systems_1993}.

We average observables over 200--1000 trajectories. We represent the infinite-temperature initial state by sampling Haar-random pure states of the chain and initialize the auxiliary in $|{\downarrow}\rangle$. Simulations in \autoref{fig:2} use DanceQ~\cite{schaefer_danceq_high_2025, schaefer_codebase_release_2025}.

The FGR rate-theory curves are obtained independently. We diagonalize
\(\Heff\) for \(L=12\), construct the unitary
\(K(t)\) up to order
\(1/\omega^2\), and \(\Heff\) up to order \(1/\omega\). The leading
residual time-dependent term \(V_{\rm res}(t)\) is then evaluated as the term at order \(1/\omega^2\), and the Fourier components \(O^{(m)}\) entering the
auxiliary rate are computed up to order $1/\omega^2$ from
\(O_{\rm rot}(t)=K(t)OK^\dagger(t)\). We evaluate the matrix elements in
\autoref{eq:em_aux_full}--\autoref{eq:W_jit}, compute the corresponding
FGR transition rates, and integrate the rate equation for the populations \(p_\alpha(t)\) in \autoref{eq:pauli}. For the intrinsic rate in \autoref{eq:em_int}, we replace the delta function by a Gaussian of width five times the mean level spacing. The effective energy is then
\(E_{\rm eff}(t)=\sum_\alpha p_\alpha(t)E_\alpha\). No parameters are
fitted.

\paragraph{System parameters.}
In~\figref{fig:1}{c}, we vary the heating rates by changing the jitter strength $\varepsilon$. We use $\omega = 8$, $\Geff = 1.5$, $\Delta = \Lambda$, $g = 4\kappa = \nu$ and $\varepsilon \in \{0.075, 0.106, 0.130, 0.150, 0.167, 0.183, 0.198, 0.212\}$ (approximately) for the curves from the bottom to the top. These are chosen so that the heating rates arising from period jitter $\sim \varepsilon^2$ are spaced linearly.

In~\figref{fig:2}{e}, we pick the $\varepsilon$ values at different $\Geff$ values as follows. The FGR heating rate due to period jitter, evaluated at the ground state, is approximately fixed as per~\autoref{eq:W_jit}. For $\Geff \in \{0.5, 0.75, 1.0, 1.25, 1.5, 1.75, 2.0\}$, we pick $\varepsilon \in \{0.124, 0.114, 0.133, 0.147, 0.149, 0.147, 0.143\}$ (approximately) respectively. \figref{fig:2}{a,b} use $\varepsilon \approx 0.149$ and \figref{fig:2}{c,d} use $\varepsilon \approx 0.114$.
\end{document}


\title{Supplemental Material for:\\ Dissipative Stabilization of Floquet-Engineered Many-Body Order}

\author{Shreyas Raman \orcidlink{0000-0001-6640-9704}}
\thanks{equal contribution; \href{mailto:shreyasr@bu.edu}     {shreyasr@bu.edu}}
\affiliation{Department of Physics, Boston University, Boston, Massachusetts 02215, USA}
\author{Robin Sch\"afer \orcidlink{0000-0001-9728-2371}}
\thanks{equal contribution; \href{mailto:robin_schaefer@fas.harvard.edu}     {robin\_schaefer@fas.harvard.edu}}
\affiliation{Department of Physics, Harvard University, Cambridge, MA 02138, USA}

\author{Alicia J.  Koll\'ar}
\affiliation{Department of Physics, University of Maryland, College Park, MD 20742, USA}
\affiliation{Joint Quantum Institute, NIST/University of Maryland, College Park, Maryland 20742 USA}
\affiliation{Maryland Quantum Materials Center, Department of Physics, University of Maryland, College Park, MD 20742, USA}

\author{Anushya Chandran}
\affiliation{Department of Physics, Boston University, Boston, Massachusetts 02215, USA}

\date{\today}
\maketitle


\section{Period jitter and sector selection in the long-range Ising Model}
In the case of the driven long-range Ising spin chain discussed in the main text, the auxiliary cooling processes [Eq.~(8) in the main text] do not by themselves select one symmetry-broken sector over the other.
Instead, period-two jitter causes the two ferromagnetic sectors of the Ising model to heat at different rates. The steady-state sector selection therefore arises from the imbalance in jitter-induced heating: the sector with the smaller net heating rate reaches a lower effective energy under cooling and acquires larger steady-state population. The resulting state exhibits the characteristic period-doubled magnetization response of that selected sector.

This section explains the differential heating mechanism. We first derive the effective Hamiltonian of the system in the presence of period jitter, and then apply the result to the driven long-range Ising spin chain.

\subsection{Effective Hamiltonian construction}
We use the standard high-frequency van Vleck construction~\cite{rahav_effective_hamiltonians_2003, goldman_periodically_driven_2014, bukov_universal_high_2015, abanin_effective_hamiltonians_2017, eckardt_highfrequency_approximation_2015}, written in the time-domain, which is convenient for treating period-to-period jitter later.
For a periodically driven system \(H(t)=H_0+V_{\rm dr}(t)\), \(H(t+T)=H(t)\), we define the rotating-frame state by
\begin{equation}
    \ket{\psi(t)}_{\rm rot}=K(t)\ket{\psi(t)}_{\rm lab},
    \qquad
    K(t)=e^{iS(t)} .
\end{equation}
The Hamiltonian in this frame is
\begin{equation}
    H_{\rm rot}(t)
    =
    K(t)H(t)K^\dagger(t)+i\dot K(t)K^\dagger(t).
    \label{eq:Hrot}
\end{equation}
\(S(t)\) is chosen precisely (up to a time-independent gauge choice) so that the second term cancels the explicit time dependence order by order in \(1/\omega\).  The remaining time-independent part is the effective Hamiltonian to that order.

We expand the kick operator as \(S(t)=S_1(t)/\omega+S_2(t)/\omega^2+\cdots\), with each \(S_j(t)\) periodic over one drive period. Expanding~\autoref{eq:Hrot} gives
\begin{equation}
    H_{\rm rot}
    =
    H+i[S,H]-\frac12[S,[S,H]]
    -\dot S-\frac{i}{2}[S,\dot S]+\cdots .
\end{equation}
At leading order, the time-dependent drive is removed by choosing
\begin{equation}
    \dot S_1(t)=\omega V_{\rm dr}(t)
    \label{eq:dS1dt}
\end{equation}
Integrating~\autoref{eq:dS1dt} and choosing a gauge such that average of $S_1$ over a full period is $0$, one gets 
\begin{equation}
    S_1(t) = F(t) - \overline F \quad \text{where} \quad F(t) = \omega \int_0^t V_{\rm dr}(t')dt'
    \label{eq:S1}
\end{equation}
where the overbar denotes the period average. For zero-mean drives ($\int_0^T V_{\rm dr}(t)\,dt=0$), the kick operator is periodic. Thus, the gauge choice only shifts $S_1(t)$ by a constant, leaving the first-order correction $H_1$ (defined below) gauge-invariant. With this choice, the zeroth-order effective Hamiltonian is \(H_0\).  At the next order one finds
\begin{equation}
    H_{\rm rot}(t)
    =
    H_0
    +
    \frac{i}{\omega}[S_1(t),H_0]
    +
    \frac{i}{2\omega}[S_1(t),V_{\rm dr}(t)]
    -
    \frac{\dot S_2(t)}{\omega^2}
    +O(\omega^{-2}).
\end{equation}
Among the \(O(\omega^{-1})\) terms, the period average of \(\frac{i}{2\omega}[S_1(t),V_{\rm dr}(t)]\) is kept as the first correction to the effective Hamiltonian; the remaining oscillatory piece is removed by \(S_2(t)\). Thus
\begin{equation}
    H_1=\frac{i}{2T}\int_0^T dt\, [S_1(t),V_{\rm dr}(t)],
    \qquad
    \dot S_2(t) = \omega^2 \left(\frac{i}{\omega}[S_1(t), H_0] + \frac{i}{2\omega}[S_1(t), V_{\rm dr}(t)] - \frac{H_1}{\omega} \right),
    \label{eq:H1}
\end{equation}
and the rotating-frame Hamiltonian takes the form
\begin{equation}
    H_{\rm rot}(t) = \Heff + V_{\rm res}(t),
    \qquad
    \Heff = H_0 + \frac{H_1}{\omega},
    \qquad
    V_{\rm res}(t) = O(\omega^{-2}).
    \label{eq:Heff_def}
\end{equation}
This reproduces the effective Hamiltonian \(\Heff\) of Eq.~(1) of the main text.  The terms that remain time-dependent after truncating the construction are denoted \(V_{\rm res}(t)\); they generate the intrinsic Floquet heating rates \(W^{\rm int}_{\alpha\to\beta}\) [Eq.~(9) of the main text].

\subsection{Effective Hamiltonian with period jitter}
Period jitter introduces heating in the basis of $\Heff$ and one needs to be careful because the fluctuating drive duration results in mismatched rotating frames in consecutive periods. In period \(n\), the duration is
\begin{equation}
    T_n=T(1+\zeta_n),
    \qquad
    \zeta_n\in[-\varepsilon,\varepsilon],
    \qquad
    \overline{\zeta_n}=0,
    \qquad
    \overline{\zeta_n\zeta_m}=\frac{\varepsilon^2}{3}\delta_{nm}.
\end{equation}
The drive amplitudes are held fixed; all segments in a given period are stretched by the same factor \(1+\zeta_n\).  Therefore the kick-frame construction can be repeated period by period. To linear order in \(\zeta_n\),
\begin{equation}
    H_{\rm eff}^{(n)}
    =
    H_0+\frac{1+\zeta_n}{\omega}H_1,
    \qquad
    S^{(n)}(t)
    =
    \frac{1+\zeta_n}{\omega}S_1(t)
    +
    \frac{(1+\zeta_n)^2}{\omega^2}S_2(t)+\cdots .
\end{equation}
The important new feature is that the kick frame changes between neighboring periods.  The stroboscopic evolution in the rotating frame is approximately (neglecting higher order time-dependent terms in $1/\omega$)
\begin{equation}
    U_{{\rm rot},n}
    =
    e^{iS^{(n+1)}(0)}
    e^{-iS^{(n)}(0)}
    e^{-iH_{\rm eff}^{(n)}T_n}
    \equiv
    e^{-i\mathcal H_{\rm eff}^{(n)}T}.
    \label{eq:Urot_jitter}
\end{equation}
The first two factors are the mismatch between the kick frame at the end of period \(n\) and the kick frame at the beginning of period \(n+1\).

Using the BCH expansion on~\autoref{eq:Urot_jitter} and retaining terms up to order \(\varepsilon/\omega\), the stochastic effective Hamiltonian is
\begin{equation}
    \mathcal H_{\rm eff}^{(n)}
    =
    H_{\rm eff}
    +
    \zeta_n A_{\rm jit}
    +
    \eta_n B_{\rm jit}
    +\cdots ,
    \label{eq:Hjit}
\end{equation}
with $\eta_n = \zeta_{n+1} - \zeta_n$ and
\begin{equation}
    A_{\rm jit} = H_0 + 2\frac{H_1}{\omega},
    \quad
    B_{\rm jit}
    =
    -
    \left(
    \frac{S_1(0)}{2\pi}
    +
    \frac{2S_2(0)}{2\pi \omega}
    -
    \frac{i}{2\omega}
    \left[
    S_1(0),H_0
    \right]
    \right).
    \label{eq:AB_jit}
\end{equation}
\autoref{eq:Hjit} separates two effects of jitter. $A_{\rm jit}$ arises from the rescaled duration of the $n^{\rm th}$ cycle, and $B_{\rm jit}$ arises from the mismatch of the rotating frames between consecutive periods. Although $\zeta_n$ and $\eta_n$ have zero mean, they are correlated because \(\eta_n\) contains \(\zeta_n\):
\begin{equation}
    \overline{\zeta_n\eta_m}
    =
    \frac{\varepsilon^2}{3}
    \left(\delta_{n,m+1}-\delta_{n,m}\right).
    \label{eq:zeta_eta_corr}
\end{equation}
This correlation controls the structure of the jitter matrix element in the FGR rate analysis.

When the jitter strength $\varepsilon$ is small, we can treat $A_{\rm jit}$ and $B_{\rm jit}$ as stochastic perturbations on $\Heff$. Let
\begin{equation}
\Heff|\alpha\rangle=E_\alpha|\alpha\rangle,
\qquad
\delta E_{\beta\alpha}=E_\beta-E_\alpha ,
\end{equation}
where $\delta E_{\beta\alpha}$ is the transition energy, in the notation of the main text.
During one period, the stochastic perturbation is $\zeta_n A_{\rm jit}+\eta_n B_{\rm jit}
=
\zeta_n(A_{\rm jit}-B_{\rm jit})+\zeta_{n+1}B_{\rm jit}$.
Averaging over the independent jitter variables gives the effective matrix element
\begin{equation}
\bra \beta M_{\rm jit} \ket \alpha
=
\bra \beta A_{\rm jit}\ket \alpha
+
\left(e^{-i\delta E_{\beta\alpha}T}-1\right)
\bra \beta B_{\rm jit}\ket \alpha.
\end{equation}
The corresponding transition rate is
\begin{equation}
W^{\rm noise}_{\alpha\to\beta}
=
\frac{4\varepsilon^2}{3T}
\left|\bra \beta M_{\rm jit}\ket \alpha\right|^2
\frac{\sin^2(\delta E_{\beta\alpha}T/2)}
{\delta E_{\beta\alpha}^2}.
\label{eq:Wnoise}
\end{equation}
This is the noise rate $W^{\rm noise}_{\alpha\to\beta}$ quoted in the End Matter [Eq.~(10) of the main text].

\subsection{Consequences for the driven long-range Ising Model}
In this section, we describe how introducing period-two jitter in the driven long-range Ising model discussed in the main text differentially heats states in the two ferromagnetic sectors. The one-period Floquet unitary is [Eq.~(6) of the main text]
\begin{equation}
    U(T)=R_xU_z(T/2)U_y(T/2),
\end{equation}
where \(R_x=\prod_i\sigma_i^x\) is a global \(\pi\)-pulse, \(U_\alpha(t)=e^{-iH_\alpha t}\), $H_\alpha = H_0 + \frac{B}{2}\sum_i \sigma_i^\alpha$, and $H_0 = -\frac{J}{4}\sum_{i>j}\frac{\sigma_i^z \sigma_j^z}{|i-j|^p}$, with $p=1.5$. Since conjugating by \(R_x\) flips the sign of \(\sigma^{y,z}\) and leaves \(H_0\) invariant, \(R_xH_\alpha R_x\equiv \tilde H_\alpha=H_0-\frac{B}{2}\sum_i \sigma_i^\alpha\). The two-period unitary is therefore [Eq.~(7) of the main text]
\begin{equation}
    U(2T)
    =
    \tilde U_z(T/2)\tilde U_y(T/2)
    U_z(T/2)U_y(T/2),
    \label{eq:two_period_tc}
\end{equation}
where \(\tilde U_\alpha(t)=e^{-i\tilde H_\alpha t}\). Following the same high-frequency construction as~\autoref{eq:S1}--~\autoref{eq:Heff_def},~\autoref{eq:two_period_tc} realizes a two-period effective Hamiltonian \(\Heff^{(2T)}\),
\begin{equation}
    \Heff^{(2T)} = H_0 - \frac{B^2 \pi}{8 \omega} \sum_i\sigma_i^x.
    \label{eq:Heff2T}
\end{equation}
$\Heff^{(2T)}$ possesses the Ising symmetry \(R_x\).  The two ferromagnetic sectors of \(\Heff^{(2T)}\) are related by \(R_x\), and the lab-frame \(\pi\) pulse swaps them after every period.

We now add period jitter to this protocol. We draw a new $\zeta_n$ uniformly from $[-\varepsilon_2,\varepsilon_2]$ for each pair of drive cycles, matching the period $2T$ of the effective Hamiltonian in whose basis we work. Here $n$ indexes pairs of cycles. We emphasize that period-two jitter is essential to differential heating. Repeating the steps above,
\begin{equation}
    \mathcal H_{\rm eff}^{(2T,n)}
    =
    \Heff^{(2T)}
    +
    \zeta_n A_{\rm jit}^{(2T)}
    +
    \eta_n B_{\rm jit}^{(2T)}
    +\cdots ,
    \qquad
    \eta_n=\zeta_{n+1}-\zeta_n ,
    \label{eq:Hjit2T}
\end{equation}
in direct analogy with~\autoref{eq:Hjit}--~\autoref{eq:AB_jit}, the operators carrying the same meaning for this drive. To linear order in $\varepsilon_2$, the jitter perturbation contains operators with the symmetry properties
\begin{equation}
    R_x A_{\rm jit}^{(2T)}R_x=A_{\rm jit}^{(2T)},
    \qquad
    R_x B_{\rm jit}^{(2T)}R_x=-B_{\rm jit}^{(2T)} .
    \label{eq:parity}
\end{equation}
For a transition \(\alpha\to\beta\) in one ferromagnetic sector, the jitter matrix element is
\begin{equation}
    \bra \beta M_{\rm jit}^{(2T)} \ket \alpha
    =
    \bra \beta A_{\rm jit}^{(2T)} \ket \alpha
    +
    \left(e^{-2i\delta E_{\beta\alpha}T}-1\right)
    \bra \beta B_{\rm jit}^{(2T)}\ket \alpha .
    \label{eq:Mjit_TC}
\end{equation}
For the symmetry-related transition $R_x \ket \alpha \to R_x \ket \beta$ in the other ferromagnetic sector,~\autoref{eq:parity} leaves the first term unchanged and flips the sign of the second.  The two sectors therefore have different jitter-induced heating rates $W_+^{\rm noise}$ and $W_-^{\rm noise}$ (for the positive- and negative-magnetization  sectors):
\begin{equation}
    W_+^{\rm noise}-W_-^{\rm noise}
    \propto
    {\rm Re}
    \left[
    \left(e^{-2i\delta E_{\beta\alpha}T}-1\right)
    \bra \alpha A_{\rm jit}^{(2T)}\ket \beta
    \bra \beta B_{\rm jit}^{(2T)}\ket \alpha
    \right].
    \label{eq:bias}
\end{equation}
This term is generically nonzero and is \(O(\varepsilon_2^2)\), the same order as the period-two-jitter-induced rate itself. The jitter-induced heating therefore distinguishes the two sectors, while the Ising symmetry of \(\Heff^{(2T)}\) remains intact.

Period-one jitter is redrawn every period, so its statistics are invariant under a shift by one period. The shift exchanges the two ferromagnetic sectors, and symmetry-related preparations have equal jitter-averaged heating rates. Period-two jitter is redrawn every other period. Its statistics are invariant under shifts by two cycles, allowing the rate difference in~\autoref{eq:bias}. To leading order the rates from the two independent jitter sources add, so period-one jitter contributes a sector-symmetric heating channel.

This rate asymmetry is the origin of the sector selection described in the main text. The auxiliary cooling operator used in the numerics is local, proportional to \(\sigma_1^y\), and is odd under the Ising symmetry \(R_x\). Nevertheless, cooling alone does not select one symmetry-broken sector over the other: the corresponding transition amplitudes between \(R_x\)-related states differ only by a sign, so the associated rates are unchanged. This is consistent with the numerical observation that cooling with period-one jitter alone produces correlations across the chain with vanishing net magnetization [\autoref{fig:supp_overview_mag}(b) and~\autoref{fig:supp_hidden}].

For the parameters of Fig.~3 in the main text, period-two jitter favors the positive-magnetization sector. The cooled state thus has positive magnetization on even cycles and negative magnetization on odd cycles. The unequal decay times without cooling are shown in~\autoref{fig:supp_updown}; with cooling, even the all-down preparation evolves toward positive even-cycle magnetization [\autoref{fig:supp_overview_mag}(d)].

Going forward, keeping the same convention as the main text, we will drop the $2T$ superscript in the effective Hamiltonian. Any reference to $\Heff$ for the Long-Range Ising model will refer to the effective Hamiltonian defined over two periods as in~\autoref{eq:Heff2T}.


\section{Numerical results for the driven long-range Ising chain}
\label{sec:tc_numerics}
\begin{figure}[t]
    \centering
    \includegraphics[width = 0.95 \linewidth]{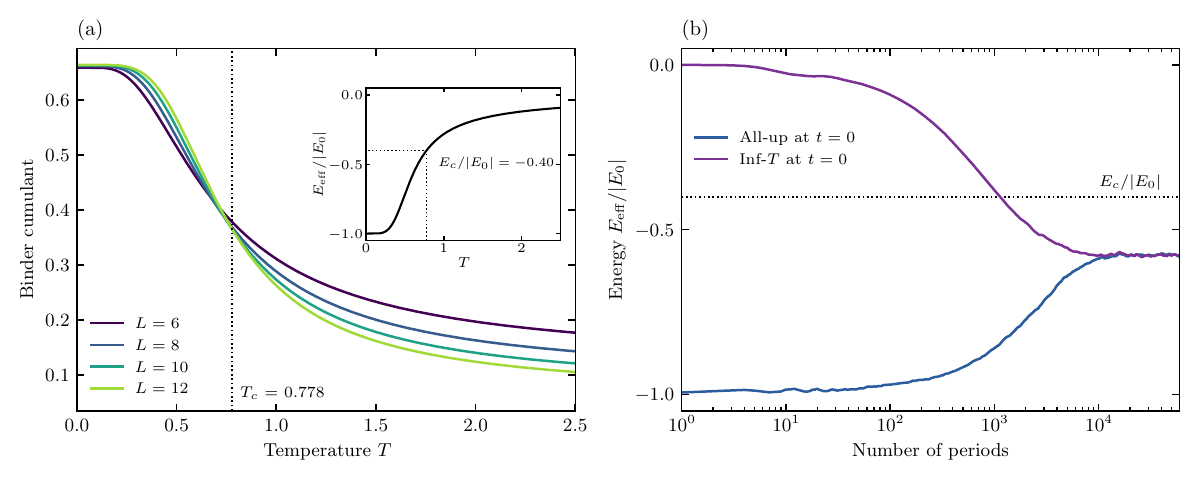}
    \caption{Critical energy of the effective long-range transverse-field Ising model. (a)~Binder cumulant \(U_4\) of the thermal state of \(\Heff\) [\autoref{eq:Heff2T}] versus temperature \(T\), for \(L=6\)--\(12\). The dotted line marks the \(L=10/12\) crossing, \(T_c\approx0.78\,J\). Inset: thermal energy \(E_{\rm eff}/|E_0|\) of the \(L=10\) chain versus \(T\); \(T_c\) maps to the critical energy \(E_c/|E_0|\approx-0.40\). (b)~Effective energy versus the number of drive periods with heating and cooling, for the polarized-up and infinite-temperature initial states. Both relax to below \(E_c\) (dotted line). Simulation parameters for (b) are $L = 12, p=1.5, \omega/J = 16, B/J = 2.257, \Delta/J = 1, \kappa/J = 0.25$, $\varepsilon_1 = 0.05$ and $g/J = 1$ when the auxiliary is coupled.}
    \label{fig:supp_critical}
\end{figure}

This section presents further evidence that
auxiliary cooling stabilizes discrete time-crystalline order in
the long-range Ising chain of Fig.~3 of the main text. Period-one
jitter heats the uncooled chain to infinite temperature
[\autoref{fig:supp_updown}(a)], while auxiliary cooling produces
a steady state below the equilibrium ordering threshold of the
two-period effective Hamiltonian [\autoref{fig:supp_critical}].
The cooled steady state retains long-range Ising correlations
[\autoref{fig:supp_hidden}], but its net magnetization vanishes
because the two ferromagnetic sectors are equally populated
[\autoref{fig:supp_overview_mag}(b)]. Weak additional period-two
jitter makes the sectors heat at different rates
[\autoref{fig:supp_updown}(b,c)]. Cooling then favors the
slower-heating sector, producing the same steady-state
period-doubled response from all three initial states
[\autoref{fig:supp_overview_mag}(d)].

\subsection{Numerical details}
\label{sec:tc_details}

We simulate chains of \(L=12\) spins with \(p=1.5\), \(J=1\), \(\omega=16\) and \(B=2.257\), so that the transverse field in~\autoref{eq:Heff2T} is \(B^2\pi/(8\omega)=0.125\). When present, the auxiliary couples to the first spin through \(O=\sigma_1^y\), with \(g=1\), \(\kappa=g/4=0.25\) and \(\Delta=1\). When the auxiliary is present, each cycle ends
with the joint pulse $P=\tau^z\otimes R_x$, which commutes with the coupling $\tau^y\sigma_1^y$; without the auxiliary, the pulse is $R_x$. Two independent jitter sources act on the drive: period-one jitter, drawn uniformly from \([-\varepsilon_1,\varepsilon_1]\) every period, and period-two jitter, drawn uniformly from \([-\varepsilon_2,\varepsilon_2]\) once per pair of consecutive periods. Each period-two draw is applied to both periods of its pair, and the first pair begins at the start of every trajectory. Jitter values are sampled independently across trajectories. The two contributions add to the fractional change in the period. Throughout, we refer to period-one jitter with \(\varepsilon_1=0.05\) as heating and to period-two jitter with \(\varepsilon_2=0.07\) as bias.

We evolve the full time-dependent Hamiltonian with quantum trajectories as described in the End Matter, using 50 time steps per period, for \(6\times10^4\) periods. The initial state of the chain is fully polarized up, fully polarized down, or a Haar-random pure state, which samples the infinite-temperature state; the auxiliary starts in \(\ket{\downarrow}\). We record the stroboscopic magnetization \(S^z_{\rm tot}/L\) in the lab frame and the effective energy \(E_{\rm eff}=\bra{\psi_{\rm rot}}\Heff\ket{\psi_{\rm rot}}\), where \(\ket{\psi_{\rm rot}}=e^{iS(0)}\ket{\psi}\) is the rotating-frame state with \(S\) constructed to order \(1/\omega^2\). We characterize the period-doubled response by the net magnetization $S^z_{\rm tot}/L$ averaged separately over
even and odd periods in the last \(2\times10^4\) periods. Because the magnetization is measured in the lab frame, the polarized state has \(S^z_{\rm tot}/L\approx-0.45\) rather than \(-0.5\) at odd periods.

We average over approximately 400 trajectories with the auxiliary and 200 without it. For display, we smooth each trajectory with a centered time window spanning approximately 5\% of the elapsed time, treating even- and odd-period magnetizations separately. Energy traces are smoothed over all periods. The 16\%--84\% bootstrap confidence intervals are narrower than the plotted lines and are not shown. Quoted steady-state averages use the unsmoothed samples from the final $2\times10^4$ periods.

\FloatBarrier

\subsection{Critical energy}
\label{sec:tc_critical}
The equilibrium ordering threshold provides a
benchmark for whether cooling can sustain ferromagnetic order.
We estimate this threshold from the Binder cumulant
\(U_4=1-\langle m^4\rangle/(3\langle m^2\rangle^2)\), with
\(m=S^z_{\rm tot}/L\), evaluated in thermal states of \(\Heff\)
[\autoref{fig:supp_critical}(a)]. The crossing of the two largest
system sizes defines an estimate of the critical temperature.
We obtain the corresponding energy \(E_c\) from the equilibrium
energy of the \(L=10\) chain. With period-one jitter and cooling, the steady state effective energy lies below \(E_c\)
[\autoref{fig:supp_critical}(b)]. The auxiliary therefore maintains the chain below the equilibrium ordering threshold despite continued heating.

\subsection{Differential heating of the two sectors}
\label{sec:tc_differential}
\begin{figure}[!htbp]
    \centering
    \includegraphics[width=\linewidth]{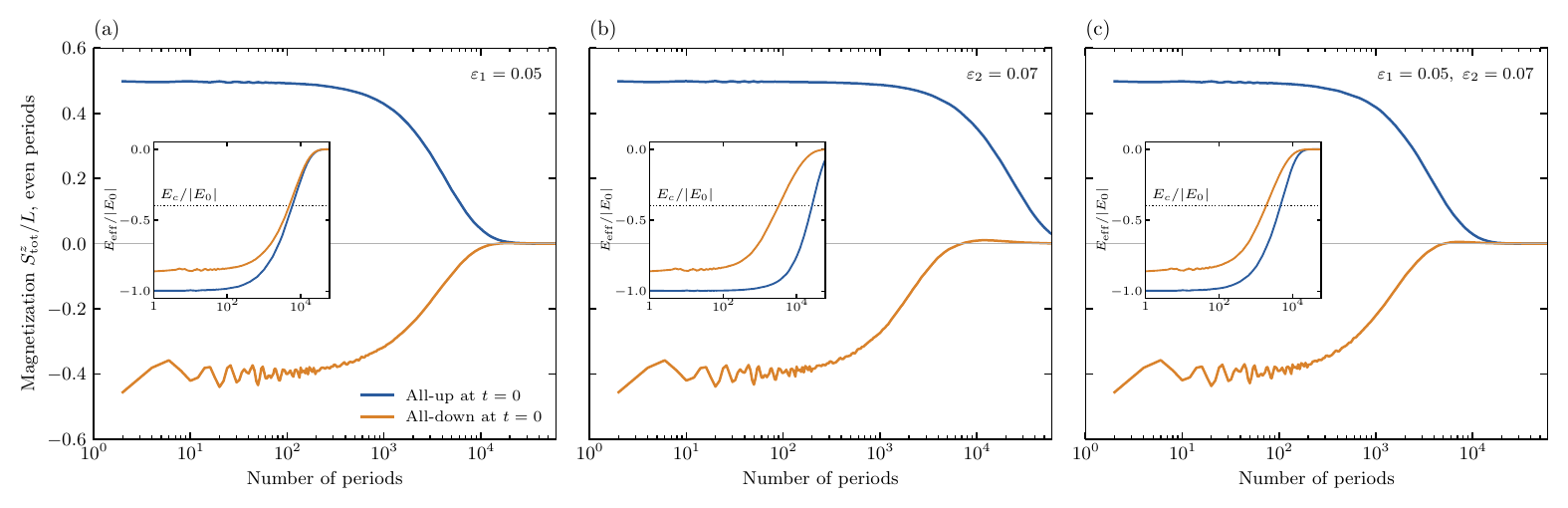}
    \caption{Differential heating of the two sectors without the auxiliary. Even-period magnetization \(S^z_{\rm tot}/L\) from the polarized-up and polarized-down states, for (a)~period-one jitter only (\(\varepsilon_1=0.05\)), (b)~period-two jitter only (\(\varepsilon_2=0.07\)), and (c)~both. The two initial states use the same jitter realizations. Insets: effective energy \(E_{\rm eff}/|E_0|\) of the same runs; the dotted line marks \(E_c\). The all-down preparation starts at a higher rotating-frame energy. Parameters are $L=12$, $p=1.5$, $\omega/J=16$, and $B/J=2.257$.}
    \label{fig:supp_updown}
\end{figure}

Period-two jitter selects a ferromagnetic sector
by producing unequal heating rates. To isolate this effect,
we compare the polarized-up and polarized-down preparations
without the auxiliary, using the same jitter realizations
[\autoref{fig:supp_updown}]. Period-two
jitter produces a substantial difference, with the
positive-magnetization sector heating more slowly. This is the
sector favored in the cooled steady state
[\autoref{fig:supp_overview_mag}(d)], in agreement with~\autoref{eq:bias}. Period-two jitter also causes heating, although
less strongly than period-one jitter at the strengths considered.

\FloatBarrier

\subsection{Dynamics with heating, bias and cooling}
\label{sec:tc_overview}
\begin{figure}[!htbp]
    \centering
    \includegraphics[width=\linewidth]{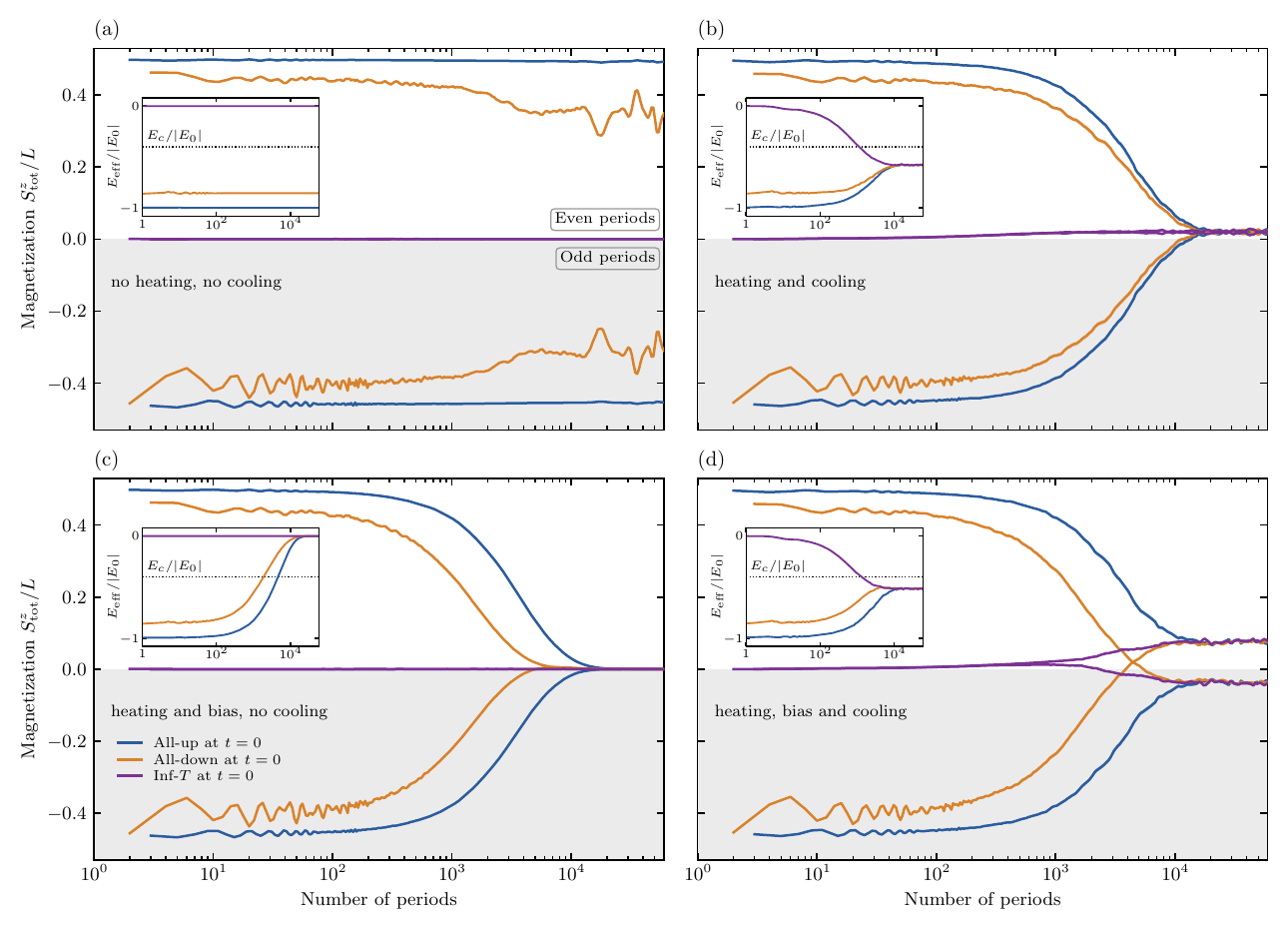}
    \caption{Stroboscopic magnetization \(S^z_{\rm tot}/L\) at even and odd periods for the polarized-up, polarized-down and infinite-temperature initial states. (a)~No jitter and no auxiliary. (b)~Heating and cooling. (c)~Heating and bias, without the auxiliary. (d)~Heating, bias and cooling. Heating denotes period-one jitter with \(\varepsilon_1=0.05\), and bias denotes period-two jitter with \(\varepsilon_2=0.07\). For the all-up initial state, the even-cycle trace starts at positive magnetization; the all-down initial state has the opposite sign. The background shading separates positive and negative magnetization. Insets: effective energy \(E_{\rm eff}/|E_0|\) of the same runs; the dotted line marks the critical energy  [\autoref{fig:supp_critical}]. In (a), the polarized states are single deterministic trajectories, and the magnetization of the polarized-down state decreases slowly by dephasing at constant energy. In (b), all initial states reach the same energy below \(E_c\) with no net period-doubled response. In (d), all initial states reach the same steady state with a finite period-doubled response; the polarized-down state at even periods crosses over to the positive-magnetization sector. Parameters: $L = 12, p=1.5, \omega/J = 16, B/J = 2.257, \Delta/J = 1, \kappa/J = 0.25$ and $g/J = 1$ when the auxiliary is coupled.}
    \label{fig:supp_overview_mag}
\end{figure}

Cooling must both counteract heating and prepare
the ordered state independently of the initial polarization.
Without jitter or cooling, the polarized-up state retains a
period-doubled response over the simulated time, while the
infinite-temperature state shows none
[\autoref{fig:supp_overview_mag}(a)]. The polarized-down
magnetization decreases through dephasing at constant effective
energy; no intrinsic heating is resolved over these times.
With heating and bias but no auxiliary, the chain instead
approaches infinite temperature and the magnetization vanishes
[\autoref{fig:supp_overview_mag}(c)].

Auxiliary cooling brings all three initial states
to the same effective energy below \(E_c\). With period-one jitter
alone, the net magnetization vanishes, although ferromagnetic
correlations survive across the chain
[\autoref{fig:supp_overview_mag}(b) and~\autoref{fig:supp_hidden}]. Adding period-two jitter favors the
slower-heating sector and produces a steady-state period-doubled
response [\autoref{fig:supp_overview_mag}(d)]. Even the
polarized-down preparation evolves toward positive magnetization
on even periods. Convergence from both polarized preparations
and the infinite-temperature state shows that cooling stabilizes the response.

\FloatBarrier

\subsection{Ferromagnetic correlations with period-one jitter}
\label{sec:hidden_order}
\begin{figure}[!htbp]
    \centering
    \includegraphics[width=0.60\linewidth]{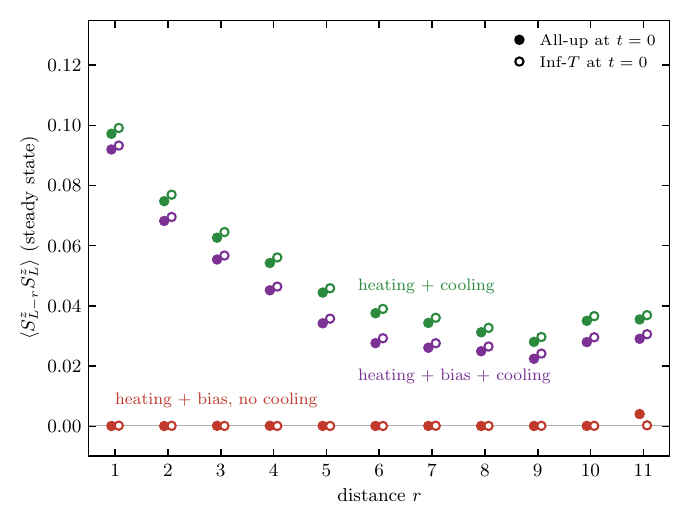}
    \caption{Ferromagnetic correlations across the finite chain. Steady-state correlator \(\langle S^z_{L-r}S^z_L\rangle\) at even periods, averaged over the last \(2\times10^4\) periods, versus distance \(r\), for heating and cooling (green), heating, bias and cooling (purple), and heating and bias without the auxiliary (red). Filled (open) symbols denote the polarized-up (infinite-temperature) initial state; the two are offset slightly in \(r\) for visibility. Parameters: $L = 12, p=1.5, \omega/J = 16, B/J = 2.257, \Delta/J = 1, \kappa/J = 0.25$ and $g/J = 1$ when the auxiliary is coupled.}
    \label{fig:supp_hidden}
\end{figure}

The vanishing net magnetization with period-one
jitter does not imply the absence of ferromagnetic order. Equally populated sectors have opposite magnetizations but retain Ising correlations. The late-time correlator
\(\langle S^z_{L-r}S^z_L\rangle\), evaluated at even periods,
remains appreciable across the cooled chain for both initial
states [\autoref{fig:supp_hidden}]. Without the auxiliary,
these correlations vanish. Cooling therefore stabilizes
ferromagnetic correlations even without the period-two bias.
With bias, the correlations are slightly weaker because the steady-state effective energy is marginally higher, but correlations still survive across the chain.

\FloatBarrier

\section{Scaling of the cooling rate with the number of auxiliaries}
\label{sec:supp_two_aux}
An assumption in the rate theory is that independent auxiliaries contribute additively to the cooling rate.  We test this directly by comparing the dynamics with one or two auxiliaries.  The auxiliaries have identical parameters \((g,\Delta,\kappa)\), and each couples locally to the same type of system operator, but at different locations in the system. In the weak-coupling regime, tracing out the auxiliaries gives a sum of independent FGR rates. Thus, the cooling rate is expected to scale linearly with the number of auxiliaries, while the intrinsic heating rate and the heating rate due to noise remain unchanged.
Concretely, near the ground state the effective energy (neglecting suppressed processes) approximately obeys the additive balance relation
\begin{equation}
    \dot E_{\rm eff}
    \;\sim\;
    -\,N_{\rm aux}\,\frac{g^2}{\kappa}\,n_{\rm qp}\,\Lambda
    \;+\;
    L\varepsilon^2 \frac{\Lambda^3}{\omega},
    \label{eq:supp_multi_aux_balance}
\end{equation}
where \(n_{\rm qp}\sim (E_{\rm eff}-E_0)/(\Lambda L)\) is the quasiparticle density and \(\Lambda\) is the center of the low-energy excitation band. The cooling term scales with the number of auxiliaries \(N_{\rm aux}\) and the heating term with the system size \(L\). At fixed heating rate due to jitter, doubling \(N_{\rm aux}\) should double the cooling rate and lower the steady-state energy. Conversely, at a fixed cooling rate due to the auxiliary, scaling $\varepsilon \to \sqrt{2} \varepsilon$ doubles the heating rate and increases the steady-state energy. Setting~\autoref{eq:supp_multi_aux_balance} to $0$, one sees that the steady state energy is controlled by the ratio $n_{\rm aux}/\varepsilon^2$, when all other parameters are held fixed.

\begin{figure}[!htbp]
    \centering
    \includegraphics[width=0.75\linewidth]{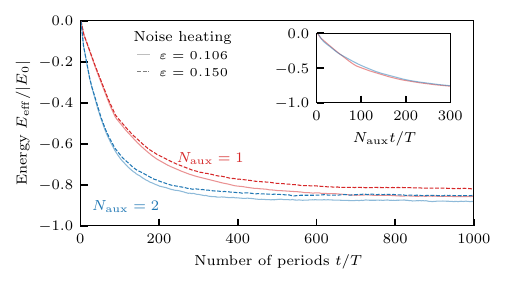}
    \caption{
    Cooling with one and two auxiliaries.  The effective energy \(E_{\rm eff}/|E_0|\) versus the number of drive periods, for \(N_{\rm aux}=1\) (red) and \(N_{\rm aux}=2\) (blue) on a chain of \(L=14\) spins.  For \(N_{\rm aux}=1\) the auxiliary sits at one end; for \(N_{\rm aux}=2\) the second auxiliary is at the center. Dashed curves with jitter strength $\varepsilon = 0.150$ have twice the noise-induced heating rate as the solid curves with $\varepsilon \approx 0.106$. Adding a second auxiliary approximately doubles the cooling rate. Inset: the same data for $\varepsilon \approx 0.106$ plotted versus \(N_{\rm aux}t/T\), showing the approximate collapse predicted by~\autoref{eq:supp_multi_aux_balance}. Parameters: $L = 14, \omega/J = 8, \Geff = 1.5, \Delta/J = 0.75, \kappa/J = 0.25, g/J = 1.0$.}
    \label{fig:supp_two_aux}
\end{figure}

\autoref{fig:supp_two_aux} confirms these expectations.  The blue curves (\(N_{\rm aux}=2\)) cool roughly twice as fast as the red curves (\(N_{\rm aux}=1\)). The solid curves correspond to $\varepsilon \approx 0.106$, while the dashed curves correspond to \(\varepsilon = \sqrt{2} \times 0.106 \approx 0.150\). Consistent with our expectations, the steady-state energy at \(N_{\rm aux}=2\) with \(\varepsilon \approx 0.150\) coincides with that at \(N_{\rm aux}=1\) and \(\varepsilon \approx 0.106\). The inset shows the same data plotted versus \(N_{\rm aux}t/T\) at a fixed $\varepsilon \approx 0.106$; the dynamics approximately collapse, as expected from \autoref{eq:supp_multi_aux_balance}.
\FloatBarrier
\section{Optimal auxiliary coupling strength}
\label{sec:supp_optimal_g}
The detuning and linewidth sweeps in Fig.~2(f) of the main text show that cooling is optimized when the auxiliary is resonant with the single-quasiparticle band of the effective TFIM. Here we perform the complementary one-dimensional sweep over the system--auxiliary coupling strength \(g\). We keep the effective transverse field at the same point in the phase diagram as in Fig.~2(f), fix the auxiliary splitting to the optimal value identified there, \(\Delta=\Lambda\), and vary \(g\) for several values of the auxiliary decay rate \(\kappa\).

For the effective TFIM,
\begin{equation}
    \Heff
    =
    -\frac{J}{4}\sum_i \sigma_i^z\sigma_{i+1}^z
    -
    \frac{J\Gamma_{\rm eff}}{4}\sum_i\sigma_i^x ,
\end{equation}
the low-energy quasiparticle band is centered at
\(
    \Lambda=\frac{J}{2}\max(1,\Gamma_{\rm eff}),
\)
with bandwidth
\(
    \nu=J\min(1,\Gamma_{\rm eff}) ,
\)
as in the main text.

We use the following conventions. The auxiliary dissipates at a rate $\kappa$, or in other words, the excited-state population decays as
\(e^{-\kappa t}\). From Eq.~(8) in the main text, the
auxiliary spectral function is a Lorentzian with \(\kappa\) as its full width at half maximum. The coupling parameter \(g\) is the coefficient appearing in the interaction
\(H_{c}=\frac{g}{4}\tau^y O^{(0)}.
\)
For a single transition \(\ket{\alpha}\to\ket{\beta}\), where $\ket \alpha, \ket \beta$ are eigenstates of $\Heff$, the matrix element is
\(V_{\beta\alpha} = \frac{g}{4} \matrixel{\beta}{O^{(0)}}{\alpha}\). Finally, \(\nu\) denotes the full peak-to-peak width of the low-energy quasiparticle band.

The physical requirement is that:
\begin{itemize}
    \item $\Delta$ is tuned to target the low-lying excitations of $\Heff$,
    \item the auxiliary dissipation rate be comparable to the bandwidth of excitations, but remains small enough to suppress off-resonant upward transitions, and
    \item the coupling matrix element is smaller than or comparable to the auxiliary dissipation rate. If $g$ gets too large, the auxiliary hybridizes appreciably with the spin that it is coupled to and can host bound states that only weakly hybridize with the quasiparticles in the rest of the system.
\end{itemize}
These considerations motivate the operating point used in the main text,
\(\Delta=\Lambda, g\simeq\nu,\kappa\simeq\nu/4
\).

\begin{figure}[!htbp]
    \centering
    \includegraphics[width=0.72\linewidth]{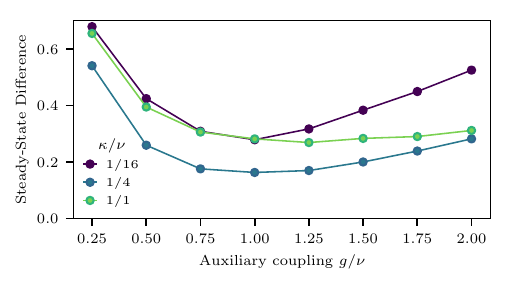}
    \caption{
Optimal auxiliary coupling strength.  Normalized steady-state energy difference
\(\frac{E_{\rm ss}-E_0}{|E_0|}\) after \(12000\) periods for \(L=12\), versus \(g/\nu\), at the same \(\Gamma_{\rm eff}\) as Fig.~2(f) of the
main text.  The auxiliary splitting is fixed at the optimal value \(\Delta=\Lambda\) from the main text, and
the three curves correspond to linewidths \(\kappa/\nu = 1/16,\,1/4,\,1\). The broad minima occur near \(g=\nu\), with the lowest energy at \(\kappa=\nu/4\).
The main-text operating point is thus chosen to be \(4\kappa=g=\nu\).  The broad minimum shows that the scheme does not require fine tuning, consistent with the design rules [Eq.~(3) of the main text]. Parameters: $L = 12, \omega/J = 8, \Geff = 1.5, \Delta/J = 0.75$ and $\nu/J = 1.0$.
}
    \label{fig:supp_optimal_g}
\end{figure}

\autoref{fig:supp_optimal_g} shows this behavior for three linewidths. For all three values of \(\kappa\), the steady-state energy first decreases as \(g\) is increased, reaches a broad minimum near \(g\simeq\nu\), and then rises again at stronger coupling. Thus, in the full many-body dynamics, the location of the optimal bare coupling is set primarily by the bandwidth of the targeted excitation band rather than by \(\kappa\) alone. The linewidth \(\kappa\) mainly controls the depth of the minimum. The best cooling occurs at the intermediate linewidth \(\kappa=\nu/4\), for which the numerical optimum \(g\simeq\nu\) also satisfies \(g\simeq4\kappa\). In this sense, the optimal parameters simultaneously target the excitation band, use the largest useful coupling strength, and match it to the auxiliary linewidth \(\kappa\). The broadness of the minimum indicates that this matching is only an order-of-magnitude design rule, not a fine-tuning condition.

\renewcommand{\selectlanguage}[1]{} 
\bibliography{ref}